\documentstyle[12pt]{article}
\def\fr{\frac}
\def\a{\alpha}

\def\l{\lambda}

\def\r{\rho}

\def\th{\theta}

\def\D{\Delta}
\def\G{\Gamma}

\def\ra{\rangle}
\def\la{\langle}
\def\be{\begin{equation}}
\def\ee{\end{equation}}
\def\bea{\begin{eqnarray}}
\def\eea{\end{eqnarray}}
\begin{document}
\begin{center}
{\large{\bf Coherent States: A General Approach}}
\end{center}
\begin{center}
P. K. Panigrahi\footnote{talk presented at the conference {\it Frontiers of 
Fundamental Physics}, held at B.M. Birla Science Center, Hyderabad,
January 2003. To appear in the proceedings.} $^{a,b}$, T. Shreecharan $^{a,b}$, 
J. Banerji $^a$ and V. Sundaram $^c$
\end{center}
\vskip0.25cm
\begin{center}
$^a$ Physical Research Laboratory, Navrangpura, Ahmedabad-380 009, India\\
$^b$ School of Physics, University of Hyderabad, Hyderabad-500 046, India\\
$^c$ Department of Electrical Engineering, IIT Madras, Chennai-600 036, India
\end{center}
\begin{abstract}
A general procedure for constructing coherent states, which are eigenstates 
of annihilation operators, related to quantum mechanical potential problems, is 
presented. These coherent states, by construction 
are not potential specific and rely on the properties of the orthogonal 
polynomials, for their derivation. The information about a given quantum
mechanical potential enters into these states, through the orthogonal 
polynomials associated with it and also through its ground state wave function. 
The time evolution of some of these states exhibit fractional revivals, 
having relevance to the factorization problem.
\end{abstract}

\section{Introduction}

Coherent states (CS) were first introduced by Schr{\"o}dinger
\cite{schro}, in the context of the harmonic oscillator, 
in an attempt to search for quantum states, whose 
time evolution followed classical equations of motion. These states were 
later studied exhaustively and expanded, because of their relevance to lasers
and various other physical problems \cite{glauber,klauder2}. 
The fact that the underlying algebraic
structure of the harmonic oscillator  and the
single mode radiation field is the Heisenberg-Weyl algebra, naturally explains
the relevance of the coherent state constructed by Schr{\"o}dinger to lasers. 
The generalizations based on $su(1,1)$ and $su(2)$ algebras followed keeping the
two mode radiation field in mind \cite{gilmore1,gsa}.  

CS are broadly separated into the following categories. The first type are
the eigenstates of certain annihilation operators. The oscillator CS belong 
to this category, as also the so-called Barut-Girardello CS \cite{barut}, 
which are the eigenstates of the annihilation operator of the $su(1,1)$ algebra. 
Based on the dynamical
symmetry possessed by the physical system, a unitary
operator $\hat U$ can be
defined such that, its action on the ground state $\mid 0 \ra$ yields a class of
states, known as the Perelomov CS 
\cite{perelomov}. States having minimum uncertainty product 
e.g., $\D x\D p = 1/2$ (in the unit $\hbar$=1) for the oscillator, 
are known as the minimum 
uncertainty CS. These type of CS have been generalized to a wide class of one 
dimensional exactly solvable potentials by Nieto and Simmons \cite{nieto1}. 
The maintainance of temporal stability under time evolution 
have recently led to a new class of CS \cite{klauder1}. As is well-known, 
for the
harmonic oscillator all these definitions lead to the same coherent state.

CS possess many interesting properties and have been
extensively used to explain a wide range of physical phenomena, see
\cite{gilmore1} for an excellent review. In general, CS are not orthogonal 
to each other and provide an overcomplete basis set. 
Recently, there has been intense activity in the study of CS \cite{klauder3}, 
largely inspired by 
the phenomena of wave packet revivals and fractional
revivals in atomic systems \cite{banerji}.
The Rydberg atom wave packets provide one such example \cite{fox}. 
It is interesting to note that, some of these 
wave packets can be utilized for the factorization of numbers
\cite{mack}.

The phenomena of wave packet revival and fractional revival occurs
in physical systems, whose energy has nonlinear terms, e.g., $E_n \propto n^2$. 
A number of quantum mechanical systems, e.g., square-well \cite{styer}, 
Morse \cite{morse} and P{\"o}schl-Teller \cite{teller} potentials, exhibit
quadratic spectra. Wave packets related to potential wells have been studied
exhaustivley \cite{schleich}, although the same is not true for the later two
potentials. 
Hence, it is appropriate that one gains a better understanding of the CS of 
these potentials \cite{klauder3,benedict,kinani}. Here, we present a procedure
to construct these CS in a general manner.

The paper is organized as follows: In the following section CS, 
for potentials with Laguerre polynomials as
their eigenstates, modulo the measure, is constructed. 
In section 3, we derive the CS for
those exactly solvable potentials, whose eigenstates can be
expressed in terms of hypergeometric functions. These can be used to construct 
CS for P{\"o}schl-Teller potential. We present our conclusions in section 4.

\section{Coherent states for potentials of confluent hypergeometric class}

In this section, we derive the coherent states for those exactly
solvable potentials whose polynomial part of the eigenstates can be written 
in terms of
the Laguerre polynomials. These potentials include Morse, Coulomb and three 
dimensional isotropic oscillator. For this purpose, we make use of a novel 
method
of solving linear differential equations \cite{guru1}. 
This method connects the space of solutions to the space of monomials.

The novel form of the solution for the Laguerre differential equation 
\be \left[ x
\frac{d^2}{dx^2}+ (\l + 1 -x )\frac{d}{dx}+ n \right]L_n^\l(x) =0 
\ee 
is given by
\be \label{lagsol} L^{\l}_n(x) = \frac{(-1)^n}{n!}
\exp \left[- x \frac{d^2}{dx^2} - (\l+1) \frac{d}{dx}
\right]\,.\,x^n\quad. 
\ee 
As pointed out earlier, this form of the solution relates the solution
space to the space of monomials; this makes it easy to identify the
ladder operators at the level of monomials. These operators can be cast into the
ladder operators for Laguerre polynomials after a suitable similarity
transformation. In the present work we do not provide the details of such
construction, instead refer the readers to \cite{guru2}. We identify
\be 
K_+\equiv x
\quad,\quad K_- \equiv \left[x\fr{d^2}{dx^2} +
(\l+1)\fr{d}{dx}\right] \quad{\rm and}\quad K_3 \equiv
D+\fr{\l+1}{2} \quad, 
\ee 
as the raising, lowering and the diagonal operator respectively at the level 
of monomials:
\be 
K_+ x^n = x^{n+1}\,,\, K_- x^n = n(n+\l)x^{n-1}\,,\, K_3 x^n = \left[n+
\fr{\l+1}{2}\right]x^n \quad, 
\ee 
which satisfy 
\be 
[K_+,K_-]= -2K_3 \qquad [K_3,K_{\pm}] = \pm K_{\pm} \quad, 
\ee 
a $su(1,1)$ algebra. Having identified the necessary ladder operators and 
the algebra satisfied by them, we now proceed to construct the coherent state. 
Here, we construct the eigenstate of the lowering operator $K_-$. For this 
purpose, we first identify, the operator ${\tilde K}_+$ such that 
$[K_-,{\tilde K}_+]=1$, utilizing the procedure developed by Shanta 
et. al. \cite{shanta}. The explicit expression of ${\tilde K}_+$ in the 
present case is
\be
{\tilde K}_+ = \fr{1}{(D+\l)}x \quad.
\ee
Defining $U = \exp(\a{\tilde K}_+)$, and introducing an identity operator 
and acting $U^{-1}$ from left in $K_- \mid 0\ra = 0$, one gets
\be
U^{-1}K_- U U^{-1} \mid 0\ra = 0 \quad.
\ee
Since $U^{-1}K_- U = (K_- + \a)$ and $K_- \la x\mid 0 \ra=0$ yields 
$\la x\mid 0\ra=x^0$, the coherent state $\mid \a\ra$ in 
co-ordinate basis is given by
\be
\la x \mid \a\ra = N(\a)^{-1} S^{-1} e^{- \a{\tilde K}_+} x^0 \quad.
\ee
Here we have introduced $S^{-1}= \exp(-K_-)$ for the sake of future convenience;
this affects only the normalization. One then gets
\be
\la x \mid \a\ra = N(\a)^{-1}
S^{-1} \sum_{n=0}^{\infty} \fr{(-1)^n}{n!}(\a{\tilde K}_+)^n x^0 \quad,
\ee
here $N(\a)^{-1}$ is the normalization factor. Writing explicitly
\be
\la x \mid \a \ra = N(\a)^{-1} S^{-1} \sum_{n=0}^{\infty}
\fr{(-1)^n}{n!} \fr{(\a x)^n}{(D+\l+n) \cdots(D+\l+1)} \,.\, x^0 \,.
\ee
or
\be \label{co}
\la x \mid \a \ra = N(\a)^{-1} S^{-1}
\sum_{n=0}^{\infty}\fr{(-1)^n}{n!}\fr{(\a x)^n}{(\l+1)_n} \quad.
\ee
Here, $(\l+1)_n$ is the Pocchammer symbol and can be defined in terms of 
the gamma functions
\be \label{poch}
(\l+1)_n = \fr{\G(\l+n+1)}{\G(\l+1)} \quad.
\ee
Substituting Eq. (\ref{poch}) and Eq. ({\ref{lagsol}) in Eq. (\ref{co}), 
we obtain
\be
\la x \mid \a \ra = N(\a)^{-1} \G(\l+1) \sum_{n=0}^{\infty}
\fr{\a^n L_n^{\l}(x)}{\G(\l+n+1)} \quad.
\ee
It can be easily noticed that the coherent state constructed above does not 
contain, {\em a priori}, any information about the potential. This can be 
incorporated by muliplying the
groundstate wave function, specific to the potential under consideration, from
the left. The above equation can be cast into a compact form 
utilizing the generating function of the Laguerre polynomials \cite{stegun},
\be \label{cst}
\la x \mid \a \ra = N(\a)^{-1} \G(\l+1)(x\a)^{-\l/2} e^\a J_{\l}[2 (x\a)^{1/2}]
\quad.
\ee
here $J_{\l}[2 (x\a)^{1/2}]$ is the Bessel function of the first kind. Although
the procedure outlined here is new, this coherent state has been derived
earlier \cite{mohanty,jellal}.

\section{Coherent state of modified P{\"o}schl-Teller potential}

In this section, we derive the coherent state for the modified P{\"o}schl-Teller
potential following the technique developed in the previous section. For this
purpose, we first start with the hypergeometric differential equation and write
its solution in the form which can be used easily for constructing the coherent
state. The hypergeometric differential equation is
\be
\left[z^2 \frac{d^2}{dz^2} + (a+b+1)z \frac{d}{dz} 
+ a b - z \frac{d^2}{dz^2} - c \frac{d}{dz} \right] F(a,b;c;z)= 0 \quad, 
\ee
whose solution can be written in the novel form
\be \label{hg}
F(a,b;c;z)= (-1)^{-a}
\fr{\G(b-a)\G(c)}{\G(c-a)\G(b)}
\exp\left[\fr{-1}{(D+b)}\left(z\fr{d^2}{dz^2}+c\fr{d}{dz}\right)\right]
\, . \, z^{-a} \quad.
\ee
The above form makes easy the identification of the underlying algebraic
structure. Defining the operators
\be
K_- = \fr{1}{(D+b)}\left(z\fr{d^2}{dz^2}+c\fr{d}{dz}\right) \quad{\rm and}
\quad{\tilde K}_+ = \fr{(D+b-1)}{(D+c-1)} z
\ee
such that they satisfy the algebra $[K_-,{\tilde K}_+]=1$. 
We obtain ${\tilde U}\equiv\exp(q{\tilde K}_+)$, which will be useful for
finding the eigenstate of $K_-$ operator. Starting from 
$K_- \mid 0 \ra=0$, and proceeding, as shown earlier for the case of Laguerre
polynomials, we get 
\be
{\tilde U}^{-1}K_- {\tilde U}{\tilde U}^{-1}\mid q \ra=0 \quad,
\quad (K_- + q){\tilde U}^{-1}\mid q \ra=0 \quad.
\ee
For casting the above into a convenient form, we define, 
${\tilde S}^{-1}\equiv\exp(-K_-)$ which yields, 
\be
(K_- + q){\tilde S}^{-1}{\tilde U}^{-1}\mid q \ra=0 \quad.
\ee
The above coherent state ${\tilde S}^{-1}{\tilde U}^{-1}\mid q \ra$ in the 
cordinate basis is given by
\bea \nonumber
\la z\mid q \ra &=& e^{-K_-}
\sum_{n=0}^{\infty}\fr{(-1)^n}{n!}q^n\left[\fr{D+b-1}{D+c-1}z\right]^n z^0
\quad{\rm or}\\ \nonumber{}\\
\la z\mid q \ra &=& e^{-K_-}\sum_{n=0}^{\infty}\fr{(-1)^n}{n!}q^n z^n
\fr{(D+b+n-1)\cdots(D+b)}{(D+c+n-1)\cdots(D+c)} z^0 \, ,
\eea
and hence,
\be
\la z\mid q \ra = \sum_{n=0}^{\infty}\fr{(- q)^n}{n!}
\fr{\G(b+n)\G(c)}{\G(c+n)\G(b)}e^{-K_-}z^n \quad.
\ee
Utilizing the expression for $F(-n,b,c;z)$, the coherent state can be expressed as
\be
\la z\mid q \ra = N(q)^{-1}\sum_{n=0}^{\infty} \fr{q^n}{n!}F(-n,b;c;z)\quad,
\ee
where $N(q)^{-1}$ is the normalization constant. 
Eigenfunctions of the modified P{\"o}schl-Teller potential \cite{quesne} are 
Gegenbauer polynomials, which is a special case of hypergeometric series with
specific parameter values: $b=n+2\r$ and $c=\r+1/2$. With the above 
substitutions the coherent state is given by
\be
\la z\mid q \ra =
N(q)^{-1}\sum_{n=0}^{\infty}\fr{\G(2\r)}{\G(2\r+n)}q^n C_n^{\r}(1-2z) \quad,
\ee
where we have used the relation
\be
F(-n,n+2\r;\r+1/2;z) = \fr{n!}{(2\r)_n}C_n^{\r}(1-2z) \quad.
\ee
The coherent state derived above can be written in a compact
form using the generating function for the Gegenbauer polynomials \cite{stegun}:
\be
\la y\mid q \ra = N(q)^{-1} \G(\r+\fr{1}{2}) \exp(q
\cos\th)\left[\fr{q}{2}\sin\th\right]^{\fr{1}{2}-\r}J_{\fr{1}{2}-\r}(q \sin\th)
\quad,
\quad.
\ee
where $(1-2z)=y=\cos\th$.
The fact that modified P{\"o}schl-Teller potential has a quadratic spectra, 
leads to revivals and fractional 
reivals during the time evolution of the above CS. This is transparent in the
auto-correlation function plotted below. It is interesting to note that the
above phenomena has recently been used for the factorization of numbers.

\begin{figure}
\begin{center}
\input epsf
\leavevmode{\epsfxsize=3.0in\epsfbox{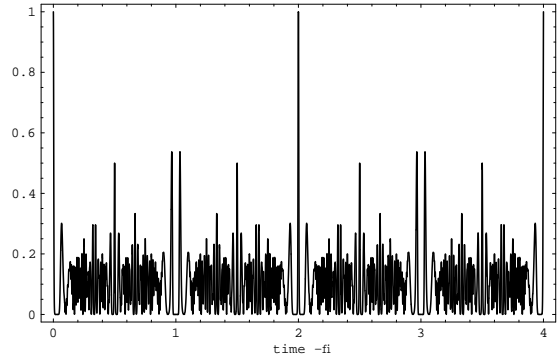}}
\end{center}
\caption{Autocorrelation function, ${\mid \la 0,y\mid q \ra \la t ,y \mid q \ra
\mid}^2$ for $\r=2$, for the coherent state (unnormalized) for
modified P{\"o}scl-Teller potential. The very large peaks denote full revival
whereas the mid peaks denote the fractional ones.}
\end{figure}

\section{Conclusions}

In conclusion, we have developed a general scheme to construct coherent states
for quantum mechanical potential problems.
The coherent states thus constructed are potential independent. The
information about the potential is incorporated by fixing the parameters of
poynomials, specific to the potential under consideration. The time evolution of
the CS leads to interesting behaviour, as manifested in the auto-correlation
function. One needs to study these aspects as also the spatio-temporal behaviour
of the CS associated with the above potential problems more carefully.


\end{document}